\documentclass[allclo]{FBSart}
\usepackage{epsfig,psfig}
\usepackage{amsfonts}

\usepackage{graphicx}

\title{Color Gauge Invariance in Hard Processes
       \footnote{Presented at Light-Cone 2004, Amsterdam, 16 - 20 August}}
\author{F.~Pijlman}
\institute{Vrije Universiteit, Department of Physics, Amsterdam}
\runningauthor{F.~Pijlman}
\runningtitle{Color gauge invariance in hard processes}
\begin{document}
\maketitle
\begin{abstract}
Within the theoretical framework we apply, a suggested origin for single spin
asymmetries is the presence of gauge links in transverse momentum dependent
distribution functions. Recently we found
new gauge link structures in a number of hard processes. These structures
need to be considered in the evolution of parton distribution functions and for
establishing factorization.
\end{abstract}

\section{Soft matrix elements in semi-inclusive DIS and Drell-Yan}

The description of the
internal structure of the nucleon is fairly well understood in terms of
transverse momentum integrated distribution functions in contrast to the
unintegrated ones. Within the formalism the latter provide access to the
transverse spin of quarks and moreover, they are sensitive to the average
gluon field in the nucleon which can be measured in single spin asymmetries
(SSA's).
Following the approach and notations of Ref.~\cite{Boer:2003}, we will 
try to understand how these distribution functions
are defined in the parton model, where it is assumed that
partons in the nucleon are soft.

\begin{figure}
\begin{center}
\begin{tabular}{lccc}
SIDIS: & \includegraphics[width=2.3cm]{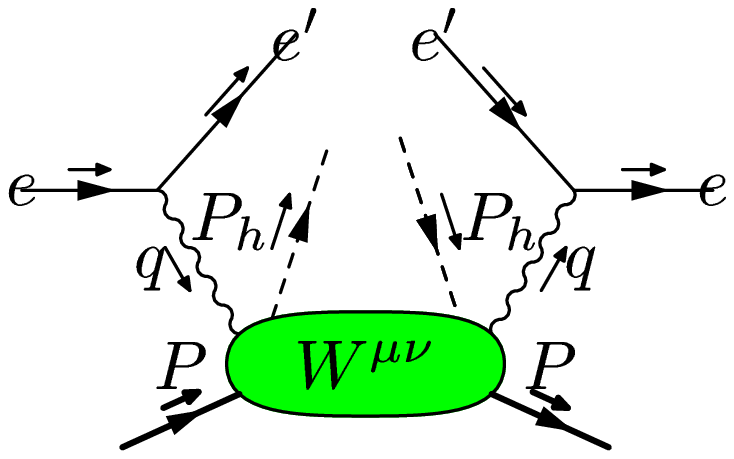}  &
\includegraphics[width=2.3cm]{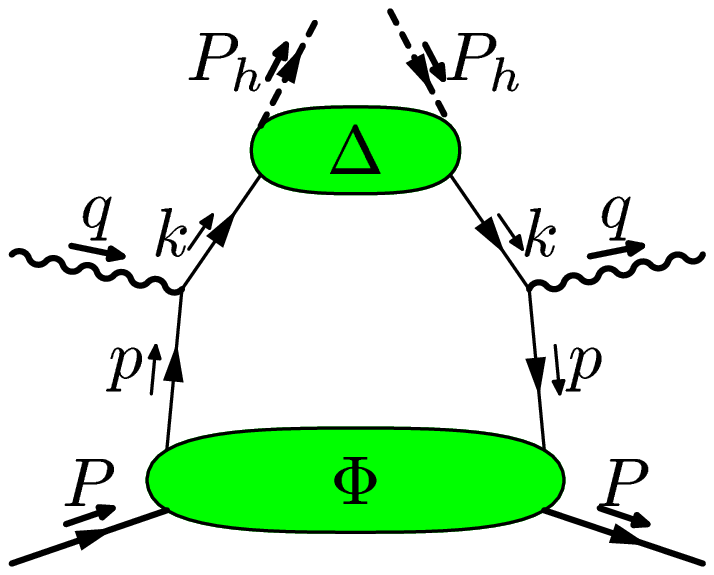} &
\includegraphics[width=2.3cm]{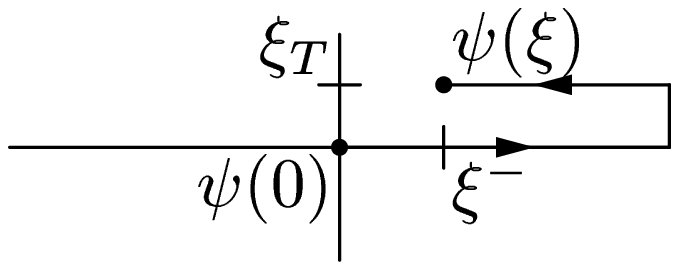}$\mathcal{L}^{[+]}$\\
DY: & \includegraphics[width=2.3cm]{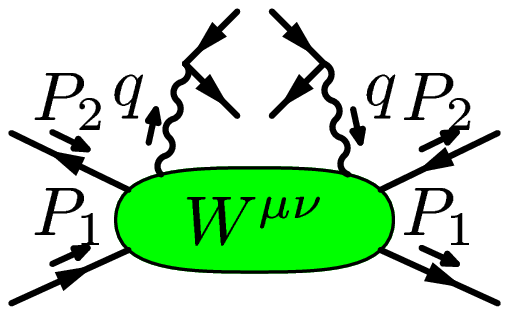} &
\includegraphics[width=2.3cm]{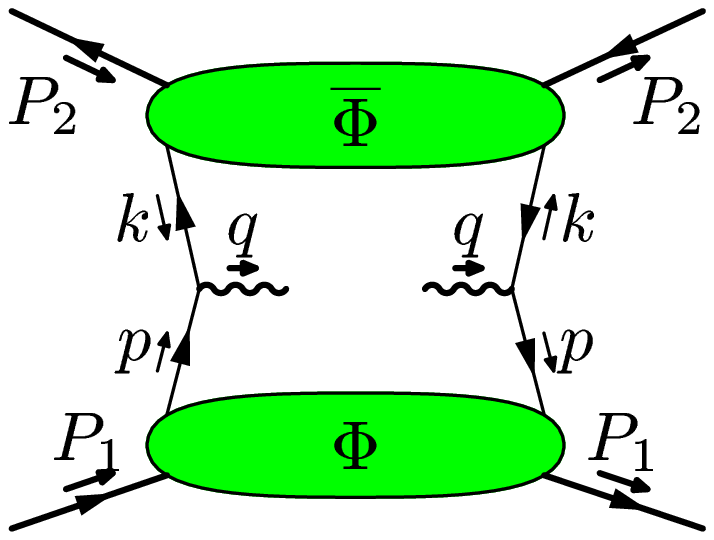}   &
\includegraphics[width=2.3cm]{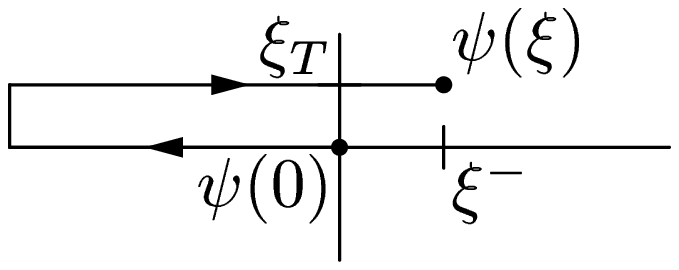}$\mathcal{L}^{[-]}$  \\
 & a & b & c
\end{tabular}
\caption{a) cross-section,
b) one of the leading contributions in the parton model, c) gauge links.
\label{processes}}
\end{center}
\end{figure}

In semi-inclusive DIS (SIDIS), see Fig.\ref{processes},
the virtual photon
couples to the current sandwiched between the in- and outgoing hadrons. The
cross-section is proportional to the
leptonic tensor, $L_{\mu\nu}$, contracted with the
hadronic tensor, $W^{\mu\nu}\sim
{}_\mathrm{in}\!\langle P | J^\mu (0) | P_X,P_h \rangle_\mathrm{out}
\ {}_\mathrm{out}\!\langle P_X,P_h|J^\nu (0) | P \rangle_\mathrm{in}$.
Assuming factorization one can show that in the parton model the virtual
photon actually
scatters off a quark for which one of the leading contributions in $M/Q$ is
given in Fig.\ref{processes}. Other
contributions at this
order involve longitudinally polarized gluons and gluon
fields at infinity. Both contributions are needed for a color gauge
invariant description,
but we will discard them for the moment. The result for the hadronic tensor,
expressed in Fourier transformed
non-perturbative matrix elements, is
\begin{eqnarray}
W^{\mu\nu} &=& \int \mathrm{d}^4 p\ \mathrm{d}^4 k\ \delta^4(p+q-k)\
\mathrm{Tr} \big[ \Phi(p) \gamma^\mu \Delta(k) \gamma^\nu \big],
\\
\Phi_{ij}(p) &=& \mathcal{FT}\ {}_\mathrm{in}\!
\langle P,S |\ \overline{\psi}_j(0)\
\psi_i (\xi)\ | P,S \rangle_\mathrm{in},\\
\Delta_{ij}(k) &=& \sum_X \mathcal{FT}\
\langle 0 | \psi_i (\xi)|  P_h,S_h;X \rangle_\mathrm{out}\ {}_\mathrm{out}\!
\langle P_h,S_h;X |
 \overline{\psi}_j(0) | 0 \rangle .
\end{eqnarray}

When considering the Drell-Yan process (DY), see Fig.\ref{processes},
the hadronic tensor in the parton model reads
\begin{eqnarray}
W^{\mu\nu} &=& \int \mathrm{d}^4 p\ \mathrm{d}^4 k\ \delta^4(p+k-q)\
\mathrm{Tr} \big[ \Phi(p) \gamma^\mu \overline{\Phi}(k)
 \gamma^\nu \big],
\\
\Phi_{ij}(p) &=& \mathcal{FT}\ {}_\mathrm{in}\! \langle P,S
 |\ \overline{\psi}_j(0)\
\psi_i (\xi)\ | P,S \rangle_\mathrm{in} .
\end{eqnarray}

Without the discarded gluon contributions we find that
the non-perturbative matrix elements
are process independent. To prove the universality of these
matrix elements, one should take gluon contributions into account and
show that the processes
factorize into a hard and a universal soft part
(the non-perturbative information). In such proofs
infrared divergences, appearing at higher order in the coupling constant,
have to be absorbed in the soft parts in a universal manner~\cite{Ellis}.
Recently this topic received a lot of attention~\cite{Ji:2004aa,Metz2}.

\section{Single spin asymmetries and gauge links}

Recent measurements~\cite{Class} with polarized electrons
and unpolarized hadrons show significant SSA's.
In this case the asymmetry is proportional to the
antisymmetric part of the leptonic tensor contracted with the hadronic
tensor. The derivation
\begin{eqnarray}
W^{\mu\nu}&\sim& {}_\mathrm{in}\!\langle P |\ J^\mu(0)\ | P_X,P_h \rangle_\mathrm{out}\ \
                          {}_\mathrm{out}\!\langle P_X,P_h |\ J^\nu(0)\ | P \rangle_\mathrm{in}
\nonumber \\
&=& {}_\mathrm{out}\!\langle P_X,P_h |\ J^{\mu\dagger}(0)\ | P \rangle_\mathrm{in}^* \ \
    {}_\mathrm{in}\!\langle P |\ J^{\nu\dagger}(0)\ | P_X,P_h \rangle_\mathrm{out}^* \nonumber\\
&=& {}_\mathrm{in}\!\langle P |\mathcal{T}^\dagger\mathcal{T} J^{\nu\dagger}(0)
               \mathcal{T}^\dagger\mathcal{T}| P_X,P_h \rangle_\mathrm{out}^*\ \
    {}_\mathrm{out}\!\langle P_X,P_h | \mathcal{T}^\dagger\mathcal{T}
                     J^{\mu\dagger}(0)
                     \mathcal{T}^\dagger\mathcal{T}| P \rangle_\mathrm{in}^*
\nonumber\\
&\neq^\mathcal{T} & {}_\mathrm{in}\!\langle \overline{P} |\ \mathcal{T} J^{\nu\dagger}(0)
        \mathcal{T}^\dagger\ | \overline{P}_X,\overline{P}_h \rangle_\mathrm{out}
\ \    {}_\mathrm{out}\!\langle \overline{P}_X,\overline{P}_h |\ \mathcal{T}
                     J^{\mu\dagger}(0)
                     \mathcal{T}^\dagger\ |\overline{P} \rangle_\mathrm{in}
\nonumber\\
&=^\mathcal{P}&
{}_\mathrm{in}\!\langle P |\ \mathcal{P}^\dagger\mathcal{T}^\dagger J^{\nu\dagger}(0)
               \mathcal{T}\mathcal{P}\ | P_X,P_h \rangle_\mathrm{out}\
    {}_\mathrm{out}\!\langle P_X,P_h |\ \mathcal{P}^\dagger\mathcal{T}^\dagger
                     J^{\mu\dagger}(0)
                     \mathcal{T}\mathcal{P}\ | P \rangle_\mathrm{in}
\nonumber\\
&\sim& W^{\nu\mu},
\end{eqnarray}
shows that the hadronic tensor can have an antisymmetric part because
under time reversal out-states transform into spatial momentum reversed
(e.g. $P \rightarrow \overline{P}$) in-states
$\mathcal{T} | P_X,P_h \rangle_{\mathrm{out}} =
| \bar{P}_X, \bar{P}_h \rangle_{\mathrm{in}}
\neq | \bar{P}_X, \bar{P}_h \rangle_{\mathrm{out}}$.
The possible mechanism of final state interactions
in quark fragmentation was suggested as an explanation in the parton
model~\cite{Collins}.

A decade later, a model calculation showed that
SSA's can be
generated without considering the fragmentation of quarks~\cite{BHS}.
This was a sign for a new source for SSA's which was
missing in the parton model description. Within the formalism outlined in
section 1, this same effect can be attributed to gluon contributions.
The longitudinally
polarized
gluons can be resummed, giving
a gauge link running from each quark field to
infinity for SIDIS~\cite{Efremov:1981}.
The fact that in DY the gauge link runs
via minus infinity was not considered to be important since in the
light-cone gauge the links would vanish. The other leading
contributions are the gluon fields at infinity and
they turned out to be the missing
part in the parton model description.
It was shown that one can resum those
contributions as well, giving a transverse link~\cite{Belitsky:2002}.
Taking all these effects
into account one finds that the leading order
expression for the hadronic tensors only differs by
 the presence of a gauge link ($\mathcal{L}$) in the matrix elements
(see Fig.\ref{processes}c)
\begin{eqnarray}
\mathrm{SIDIS}:
& &\Phi_{ij}(p) \rightarrow \Phi^{[+]}_{ij}(p) = \mathcal{FT}\
\langle P,S |\ \overline{\psi}_j (0)\ \mathcal{L}^{[+]}(0;\xi)\
\psi_i (\xi)\ | P,S \rangle,\\
\mathrm{DY}:
& &\Phi_{ij}(p) \rightarrow \Phi^{[-]}_{ij}(p) = \mathcal{FT}\
\langle P,S |\ \overline{\psi}_j (0)\ \mathcal{L}^{[-]}(0;\xi)\
\psi_i (\xi)\ | P,S \rangle.
\end{eqnarray}
One can even obtain a color gauge invariant description at
next-to-leading order in $M/Q$~\cite{Boer:2003, Boer:1999}
from which explicit asymmetries in SIDIS
were obtained~\cite{Bacchetta}.

Having \emph{different}
 but explicit gauge invariant matrix elements leads to several
interesting properties. 
For instance, assuming factorization in SIDIS and DY one can still relate
their matrix elements via a time reversal operation. It is possible to divide
the matrix elements in a process independent group (called T-even) and a 
process dependent group (called T-odd) where the latter changes sign when
comparing SIDIS and DY at this order. The last group is
proportional to the gluon field~\cite{Boer:2003},
$\scriptstyle \mathcal{FT}\
\langle P,S |\ \overline{\psi}_j (0)\  \Bigl[
\int_{-\infty}^{\infty}\! \mathrm{d}\eta^-\ \mathcal{L}(0,\eta^-)\
G^{+\alpha}(\eta^-)\ \mathcal{L}(\eta^-,\xi^-)\ \Bigr]
 \psi_i (\xi)\ | P,S \rangle \nonumber
$, and can be a source for SSA's.
For fragmentation matrix elements with different links
it is complicated to derive such a relation due to the combination of 
final state interactions and gauge links. 
More information is given in
Ref.~\cite{Boer:2003, Metz2, Metz1}.

\section{Recent developments}

\begin{figure}
\begin{center}
\begin{tabular}{ccccc}
\includegraphics[width=2cm]{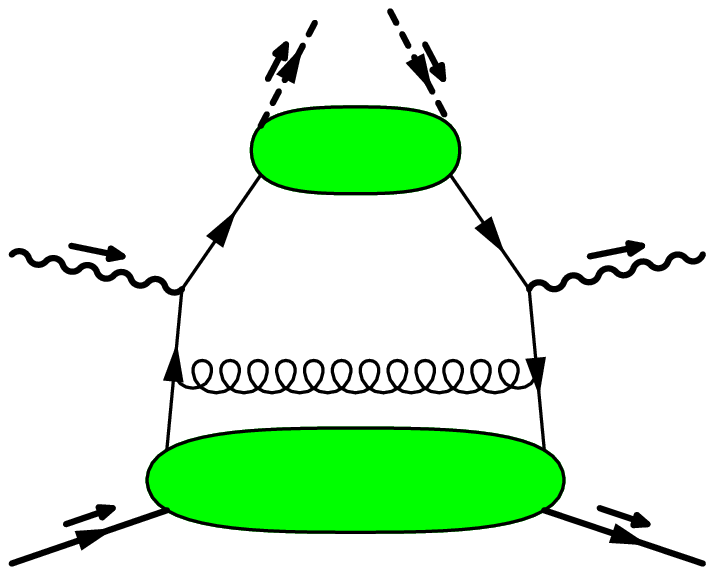}  &
\includegraphics[width=2cm]{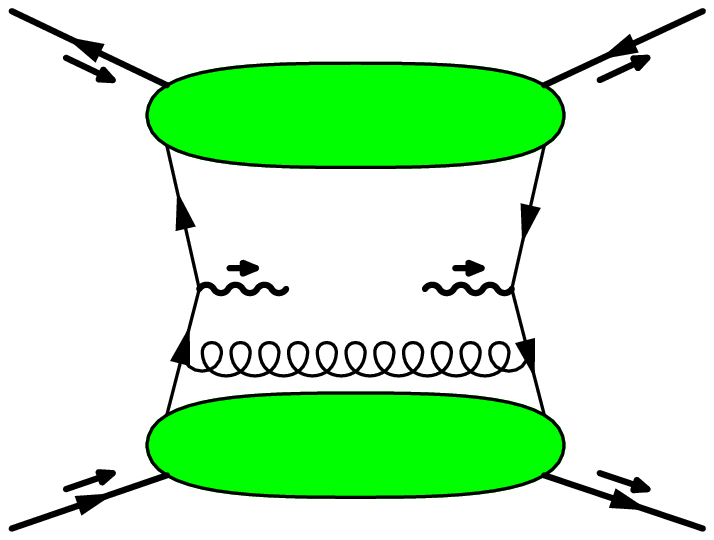}   &
\includegraphics[width=2cm]{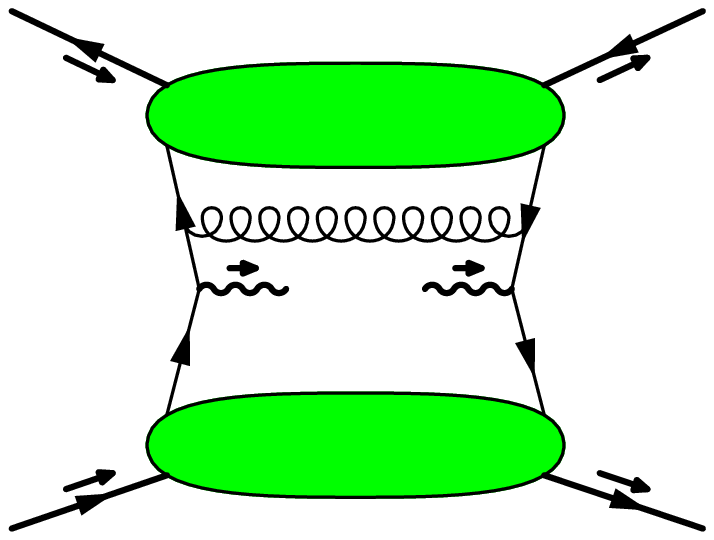} &
\includegraphics[width=2cm]{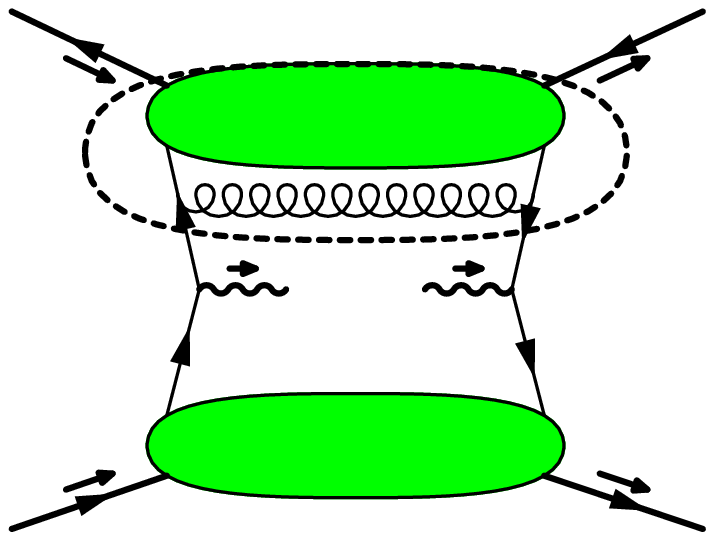} &
\includegraphics[width=2cm]{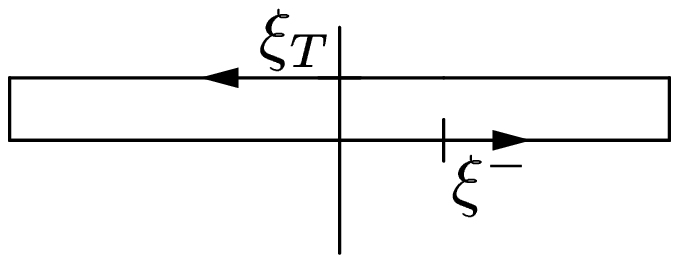} \\
a: $\scriptstyle \mathcal{L}^{[+]}$ &
b: $\scriptstyle \frac{3}{8} \mathcal{L}^{[+]}{\rm Tr}\mathcal{L}^{[ \Box ]} -
    \frac{1}{8} \mathcal{L}^{[-]}$ &
c: same as b &
d: ? &
e: $\scriptstyle \mathcal{L}^{[ \Box ]}$
\end{tabular}
\caption{The gauge links for lower blobs in several diagrams\label{Fig2}}
\end{center}
\end{figure}

Interesting effects occur when considering $\alpha_s$
corrections in SIDIS and DY.
It was shown
that closed gauge links, $\mathcal{L}^{[ \Box ]}$,
 in distribution matrix elements of a considered nucleon can
only appear
if there are \emph{other} incoming \emph{and}
outgoing QCD (bound) states~\cite{Bomhof:2004}. Therefore, such
Wilson loops do not
appear in the nucleon in SIDIS because in that case
there are, besides the nucleon, only
outgoing QCD states even when gluon bremsstrahlung is considered. 
As a result, the link still runs via plus infinity (see Fig.\ref{Fig2}a).
Supported by a first order calculation, the link very likely runs via
plus infinity for the
vertex correction and the self energy of the quark as well.
However, gluon bremsstrahlung in DY, where there are incoming hadrons, creates
outgoing QCD states.
In that case the
link
turns out to depend on the
number of gluons radiated (see Fig.\ref{Fig2}b,c).
For the
vertex and self energy correction alone, the link probably
still runs via minus infinity.

If the difference of gauge links matters (e.g. nonzero Sivers
function), then the behavior of links under
gluon radiation may imply that the
evolution of distribution functions in
DY is not the same as in SIDIS. 
This questions the universality of these functions. 
Moreover, to prove
factorization one has to absorb the infrared divergence in Fig.\ref{Fig2}d
into the upper blob. 
Since the radiated gluon also affects the link
of the lower blob, it will be difficult - if not impossible - to factorize such
diagrams.

At this moment it is unclear whether transverse momentum dependent effects
in DY can be factorized.
In
$e^+ e^-$ annihilation, where hadrons are only in the final state
and
therefore
gauge links do not change under gluon radiation,
such a factorization proof does exist~\cite{Collins1981}.
It is
expected that the incorporation of gauge links will be an essential ingredient
in considerations on factorization.

This work, done in collaboration with D.~Boer, C.~Bomhof and P.~Mulders, was
presented at a nice conference for which the organizers are acknowledged.

\end{document}